\documentclass[twocolumn]{revtex4-1}
\usepackage{graphicx}
\usepackage{graphics}
\usepackage{subfigure}
\usepackage{slashbox}
\usepackage{amsmath}
\usepackage{brackets}
\usepackage{psfrag}
\usepackage{epsfig}
\usepackage{hyperref}
\usepackage{amssymb}
\def\etal {{\it et~al.}}

\def\PRL{\rm Phys.~Rev.~Lett.~}
\def\PR{\rm Phys.~Rev.~}
\def\PRD{{\rm Phys.~Rev.} {\bf D}}

\begin{document}

\ProvideTextCommandDefault{\textonehalf}{${}^1\!/\!{}_2\ $}

\title{Search for exotic spin-dependent interactions with a spin-exchange relaxation-free magnetometer}

\author{P.-H.~Chu}
\email[Email address: ]{pchu@lanl.gov}
\author{Y.~J.~Kim}
\email[Email address: ]{youngjin@lanl.gov}
\author{I.~Savukov}
\affiliation{Los Alamos National Laboratory, Los Alamos, New Mexico 87545, USA}
\date{\today}

\begin{abstract}
We propose a novel experimental approach to explore exotic spin-dependent interactions using a spin-exchange relaxation-free (SERF) magnetometer, the most sensitive non-cryogenic magnetic-field sensor. This approach studies the interactions between optically polarized electron spins located inside a vapor cell of the SERF magnetometer and unpolarized or polarized particles of external solid-state objects. The coupling of spin-dependent interactions to the polarized electron spins of the magnetometer induces the tilt of the electron spins, which can be detected with high sensitivity by a probe laser beam similarly as an external magnetic field. We estimate that by moving unpolarized or polarized objects next to the SERF Rb vapor cell, the experimental limit to the spin-dependent interactions can be significantly improved over existing experiments, and new limits on the coupling strengths can be set in the interaction range below $10^{-2}$~m.
\end{abstract}

\maketitle

%{\it Introduction.---}
Recently, exotic spin-dependent interactions, predicted by string theories and many theoretical extensions of the Standard Model of particle physics~\cite{Antoniadis2011,Jaeckel2010,Arvanitaki2010}, have attracted much attention in the community of physicists. Such theories are associated with the spontaneous breaking of continuous symmetries, leading to massless or very light Nambu-Goldston bosons~\cite{Nambu1960,Goldstone1961,Goldstone1962}, such as the axion~\cite{Peccei1977}, and axion-like-particles (ALPs)~\cite{Jaeckel2010, Olive2014}, which are candidates for cold dark matter~\cite{Kolb1990,Bertone2005}. These exotic particles are bosons and can weakly couple with ordinary particles, such as leptons or baryons. Moody and Wilczek~\cite{Moody1984} first proposed three possible types of interactions between polarized and unpolarized particles, which were later expanded by Dobrescu and Mocioiu~\cite{Dobrescu2006} with the inclusion of the terms dependent on the relative velocity between the two interacting particles. A general classification of the interactions between particles contains sixteen types of structure of operators: fifteen of them depend on the spin of at least one of the particles and seven depend on the relative velocity of the particles. In this paper, we show a new experimental method to explore all the fifteen exotic spin-dependent interactions.
\begin{table}[b]
\begin{center}
\begin{tabular}{ c|c|c|c|c|c|c|c|c|c|c|c}
\hline
&$V_{2}$& $V_{3}$ & $V_{4+5}$ &$V_{6+7}$ &$V_{8}$ &$V_{9+10}$ &$V_{11}$ &$V_{12+13}$ &$V_{14}$ &$V_{15}$ &$V_{16}$\\
\hline
\hline
$P$ & 0 &0 &0 &0 &0 &1 &1 &1 &1 &1 &1 \\
\hline
$T$ & 0 &0 &0 &1 &0 &1 &0 &0 &1 &1 &0\\
\hline
\end{tabular}
\end{center}
\caption{The parity ($P$)- and time-reversal ($T$)-violating interactions. 1 (0) refers to the violation (no violation) of the symmetries. }
\label{tab:cpt}
\end{table}

The possible exotic spin-dependent interactions between polarized and unpolarized particles are (in SI units, adopting the
numbering scheme in~\cite{Dobrescu2006} and~\cite{Leslie2014}):
\begin{eqnarray}
&V_{4+5}  =&  -f_{4+5}\frac{\hbar^{2}}{8\pi m_{p}c}\left[\hat{\sigma}_{i}\cdot(\vec{v}\times\hat{r})\right]\left(\frac{1}{\lambda r}+\frac{1}{r^{2}}\right)e^{-r/\lambda},\label{eq:v45}
 \\
&V_{9+10}  &=  f_{9+10}\frac{\hbar^{2}}{8\pi m_{p}}(\hat{\sigma}_{i}\cdot\hat{r})\left(\frac{1}{\lambda r}+\frac{1}{r^{2}}\right)e^{-r/\lambda},
\label{eq:v910}
 \\
&V_{12+13}  &=  f_{12+13}\frac{\hbar}{8\pi}(\hat{\sigma}_{i}\cdot\vec{v})\left(\frac{1}{r}\right)e^{-r/\lambda},
\label{eq:v1213}
\end{eqnarray}	
where $m_{p}$ is the mass of the polarized particles, $\hat{\sigma}_{i}$ is the spin vector of the $i^{\text{th}}$ polarized particle while $\vec{\sigma}_{i}=\hbar\hat{\sigma}_{i}/2$, $\hbar$ is Planck's constant, $\hat{r}=\vec{r}/r$ is a unit vector in the direction between the polarized and unpolarized particles, $\vec{v}$ is their relative velocity, $c$ is the speed of light in vacuum, and $\lambda$ is the interaction range.

There are nine interactions between two polarized particles, three of which are not dependent of the relative velocity of the particles $\vec{v}$,
\begin{eqnarray}
V_{2} & = & f_{2}\frac{\hbar c}{4\pi}\left(\hat{\sigma}_1\cdot \hat{\sigma}_2\right)\left(\frac{1}{ r}\right)e^{-r/\lambda},\label{eq:v2}
\\
V_{3} & = & f_{3}\frac{\hbar^{3}}{4\pi m_{p}^2 c}[(\hat{\sigma}_1\cdot\hat{\sigma}_2)\left(\frac{1}{\lambda r^2}+\frac{1}{r^{3}}\right)\\ \notag
&-&(\hat{\sigma}_1\cdot\hat{r})(\hat{\sigma}_2\cdot\hat{r})(\frac{1}{\lambda^2 r}+\frac{3}{\lambda r^2}+\frac{3}{r^3}) ]e^{-r/\lambda},
\label{eq:v3}
\\
V_{11} & = & -f_{11}\frac{\hbar^2}{4\pi m_p}[(\hat{\sigma}_1\times\hat{\sigma}_2)\cdot\hat{r}]\left(\frac{1}{\lambda r}+\frac{1}{r^2}\right)e^{-r/\lambda},
\label{eq:v11}
\end{eqnarray}	
and the remaining six are:
\begin{eqnarray}
V_{6+7} & = & -f_{{6+7}}\frac{\hbar^{2}}{4\pi m_{p}c}\nonumber\\
&\times&\left[(\hat{\sigma}_1\cdot\vec{v})(\hat{\sigma}_2\cdot\hat{r})\right]\left(\frac{1}{\lambda r}+\frac{1}{r^{2}}\right)e^{-r/\lambda},\label{eq:v67}
 \\
V_{8} & = & f_8\frac{\hbar}{4\pi c}(\hat{\sigma}_1\cdot\vec{v})(\hat{\sigma}_2\cdot\vec{v})\left(\frac{1}{r}\right)e^{-r/\lambda},
\label{eq:8}
 \\
V_{14} & = & f_{14}\frac{\hbar}{4\pi}[(\hat{\sigma}_1\times\hat{\sigma}_2)\cdot\vec{v}]\left(\frac{1}{r}\right)e^{-r/\lambda},
\label{eq:14}
 \\
V_{15} & = & -f_{15}\frac{\hbar^3}{8\pi m_p^2 c^2 }\nonumber\\
&\times&\{[\hat{\sigma}_1\cdot(\vec{v}\times\hat{r})](\hat{\sigma}_2\cdot\hat{r})+(\hat{\sigma}_1\cdot\hat{r})[\hat{\sigma}_2\cdot(\vec{v}\times\hat{r})]\}\nonumber\\
&\times&\left(\frac{1}{\lambda^2 r}+\frac{3}{\lambda r^2}+\frac{3}{r^3}\right)e^{-r/\lambda},
\label{eq:v15}
 \\
V_{16} & = & -f_{16}\frac{\hbar^2}{8\pi m_pc^2 }\nonumber\\
&\times&\{[\hat{\sigma}_1\cdot(\vec{v}\times\hat{r})](\hat{\sigma}_2\cdot\vec{v})+(\hat{\sigma}_1\cdot\vec{v})[\hat{\sigma}_2\cdot(\vec{v}\times\hat{r})]\}\nonumber\\
&\times&\left(\frac{1}{\lambda r}+\frac{1}{\lambda r^2}\right)e^{-r/\lambda}.
\label{eq:v16}
\end{eqnarray}	
Here $f_{i}$ is the coupling strength for the interaction $V_{i}$ which can be induced from scalar, pseudoscalar, vector and axial coupling constants for the case of single massive spin-0 and spin-1~\cite{Ferreira:2015} boson exchange (see Table~1 in Ref.~\cite{Leslie2014} for details on the coupling strength). Some of these interactions are not invariant under parity-inversion ($P$) or time-reversal ($T$) symmetries. As shown in Table~\ref{tab:cpt}, the interactions $V_{11}$, $V_{12+13}$, and $V_{16}$ violate $P$ symmetry; the interaction $V_{6+7}$ violates $T$ symmetry; the interactions $V_{9+10}$, $V_{14}$ and $V_{15}$ violate both $P$ and $T$ symmetries. The $T$ and $P$-violating interactions could be induced by the axion, which is related to the strong QCD problem, or a generic light scalar boson. The detection of spin-dependent interactions, therefore, will enable to distinguish the axion from the scalar boson~\cite{Mantry:2014}.

\begin{table}
\begin{center}
\begin{tabular}{ c | c | c| c | c}
\hline
case &  $\hat{\sigma}_1$ & $\hat{\sigma}_2$ & $\vec{v}$ & interactions\\
\hline
\hline
1 & $\hat{z}$ & 0 & $\hat{z}$ & $V_{9+10}$,$V_{12+13}$\\
\hline
2 & $\hat{z}$ &  0 & $\hat{x}$ & $V_{4+5}$\\
\hline
3 & $\hat{z}$ &  0 & $\phi$ & $V_{4+5}$ \\
\hline
4 & $\hat{z}$ &  $\hat{z}$ & $\hat{z}$ & $V_2$, $V_3$,$V_{6+7}$, $V_8$\\
\hline
5 & $\hat{z}$ &  $\hat{z}$ & $\hat{x}$  &$V_{15}$\\
\hline
6 & $\hat{z}$ &  $\hat{z}$ & $\phi$& $V_{15}$\\
\hline
7 & $\hat{z}$ &  $\hat{x}$ & $\hat{z}$& $V_{11}$,$V_{16}$\\
\hline
8 & $\hat{z}$ &  $\hat{x}$ & $\hat{x}$ &$V_{16}$\\
\hline
9 & $\hat{z}$ &  $\hat{y}$ & $\hat{x}$& $V_{14}$\\
\hline
\end{tabular}
\end{center}
\caption{There are nine cases of the combination of $\hat{\sigma}_1$, $\hat{\sigma}_2$ and $\vec{v}$ for different interactions.}
\label{tab:schematic}
\end{table}

The static spin-dependent interactions have been carefully investigated for polarized electrons: experiments with a torsion pendulum~\cite{Ritter1993, Hammond2007, Heckel2008, Hoedl2011, Heckel2013, Terrano2015} and paramagnetic salt~\cite{Chui1993, Ni1994, Ni1999} and for nucleons: measurements of precession frequency of atomic gases~\cite{Vasilakis2009, Youdin1996,Chu2013,Bulatowicz2013,Tullney2013}; experiments with an ion trap~\cite{Kotler2015} and neutron bound states~\cite{Baessler2007,Jenke2014}; spin-relaxation measurements of polarized particles~\cite{Serebrov2009,Pokotilovski2010,Petukhov2010,Fu2011}. On the other hand, the experimental constraints on the interactions dependent on both spins and the relative velocity are very few. Only several experiments recently reported progress: ~\cite{Hunter2014} (measurements in the geomagnetic field),~\cite{Kimball2010} (the spin-exchange interaction studies), \cite{Piegsa2012, Yan2013} (experiments with the beam of polarized cold neutrons), and ~\cite{Yan2015} (spin relaxation studies). Furthermore, only a few new methods for polarized electrons have been proposed: the experiments with rare earth iron test masses~\cite{Leslie2014} and paramagnetic insulators~\cite{Chu2015}.

In order to probe the spin-dependent interactions for polarized electrons, we propose experimental methods based on a spin-exchange relaxation-free (SERF) magnetometer which contains 10$^{15}$ alkali atoms in a vapor cell as the source of almost 100~\% optically polarized electron spins.  The SERF magnetometer, as a type of alkali atomic magnetometer, operates in the regime of low magnetic field and high alkali density where the effect of spin-exchange collisions on spin relaxation is negligible~\cite{Happer1973,Allred:2002}. Extremely high sensitivity below 1~fT/$\sqrt{\text{Hz}}$ has been demonstrated~\cite{Kominis:2003, Sheng:2013}, which surpassed that of superconducting quantum interference devices (SQUIDs)~\cite{Kominis:2003}. Thus, the SERF magnetometer in addition to high number of polarized spins brings the advantage of very high sensitivity. Because particularly of high sensitivity, the SERF magnetometer has been employed in the test of fundamental $CPT$ symmetry~\cite{Smiciklas:2011} and in ultra-sensitive bioimaging, such as magnetoencephalography (MEG)~\cite{Xia:2006}. Recently, the SERF magnetometer has been proposed to explore the axion dark matter~\cite{Graham:2013}.

\begin{figure}[t]
\centering
\includegraphics[width=.50\textwidth]{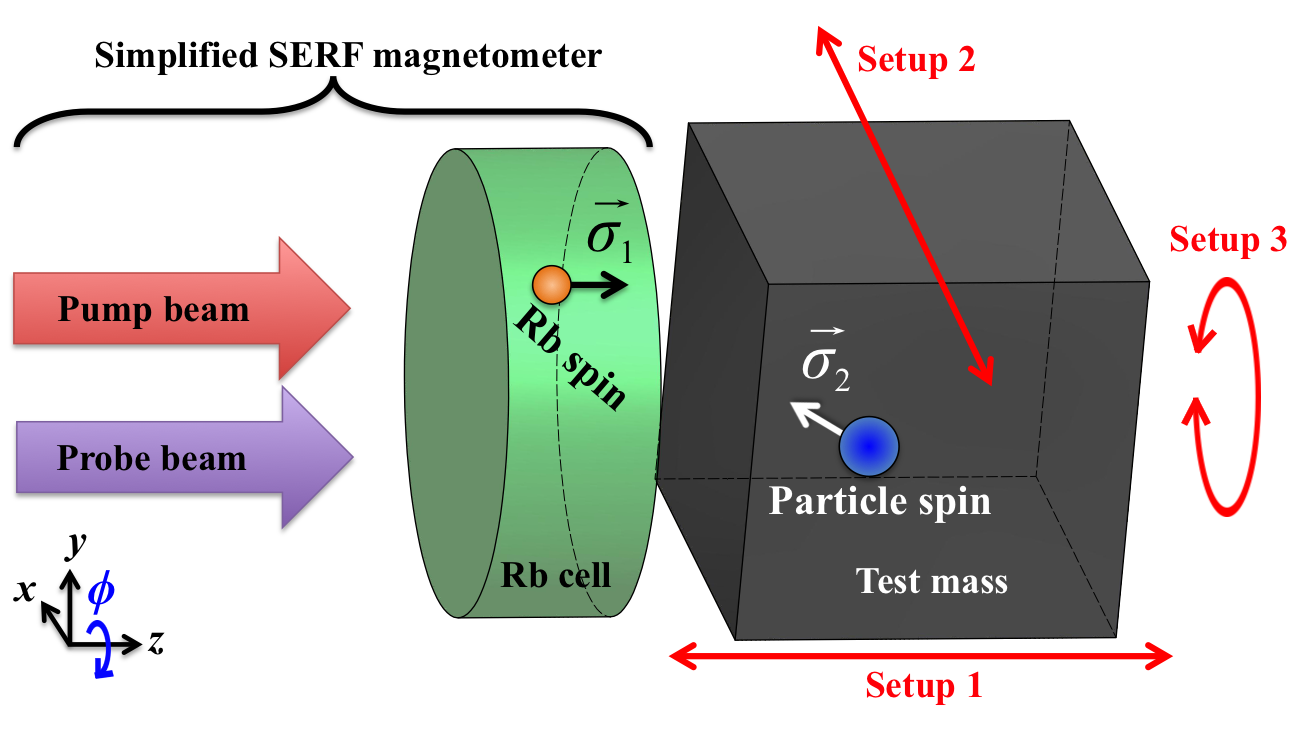}
\caption{Schematic of the experimental setup. Almost parallel pump and probe beams in the $\hat{z}$ axis are sent to a large Rb vapor cell. An unpolarized or polarized test mass is located next to the cell. The Rb electron spins in the cell and spins of the test mass are represented by $\vec{\sigma}_1$ and $\vec{\sigma}_2$, respectively. The test mass can move with a velocity $\vec{v}$ forward and backward along the $\hat{z}$ axis, move right and left along the $\hat{x}$ axis, or rotate clockwise and counterclockwise around the $\hat{z}$ axis (along the $\phi$ angle) according to the format of the exotic spin-dependent interactions (Eq.1-11). There are nine cases of the combination of $\hat{\sigma}_1$, $\hat{\sigma}_2$ and $\vec{v}$ as listed in Table~\ref{tab:schematic}.}
\label{fig:schematic}
\end{figure}

Our proposed experimental setup for the studies of the spin-dependent interactions is shown in Fig.~\ref{fig:schematic}. It is based on our SERF prototype magnetometer constructed at Los Alamos National Laboratory that demonstrated 10 fT/$\sqrt{\text{Hz}}$ sensitivity ~\cite{Kim2016}. The SERF magnetometer contains a large pancake Rb vapor cell of 6~cm diameter and 2~cm height with about 1~mm wall thickness filled with He buffer gas of 1 atm to reduce the diffusion spin relaxation. In order to generate sufficiently large Rb density, the cell is electrically heated to 150$^\circ$C. The magnetometer is placed into a magnetic field shield made of mu-metal and the residual fields inside the shield are compensated with three orthogonal coils to insure the SERF operation regime.
In a SERF magnetometer, a weak external magnetic field tilts the polarized electron spins by a small angle that depends on the magnitude of the field ~\cite{Seltzer2004, Kim2016}. The tilt is measured with a probe laser beam by its effect on the light polarization. All spin-dependent interactions have the form of $\vec{\sigma}\cdot \vec{A}$, where $\vec{A}$ is the field vector between the SERF electron spin and the interacting particle of the test mass as described by Eq.~\ref{eq:v45}-\ref{eq:v16}. The $\vec{\sigma}\cdot \vec{A}$ interactions are similar to that of an external magnetic field $\vec{B}$ with electron spin, in the form of $\vec{\sigma}\cdot\vec{B}$, and they can similarly induce the tilt of the polarized electron spins of the magnetometer. The Bloch equation, e.g. Eq.~1 in Ref.~\cite{Kim2016}, where the term of $\vec{B}$ is replaced with $\vec{A}$, can be used to describe the response of the SERF magnetometer to the spin-dependent interactions. This tilt will be measured with high sensitivity by a probe beam similarly to an external magnetic field.

The circularly polarized pump beam and linearly polarized probe beam are almost parallel and sent to the vapor cell along the $\hat{z}$ axis; thus the Rb electron spins in the vapor cell $\vec{\sigma}_1$ are pumped along the $\hat{z}$ axis. Because of this, the sensitivity of the magnetometer depends on the tilt of the spins quadratically, and to increase the magnetometer response, an offset field is applied in the transverse direction to the beams' propagation direction. The offset field can be modulated at some high frequency and the lock-in detection can be implemented to reduce the effects of laser technical noise arising from the laser frequency and intensity fluctuations which limits the magnetometer sensitivity~\cite{Kim2016}. An unpolarized or polarized cube test mass is positioned next to the cell. A mirror (not shown in Fig.~\ref{fig:schematic}) between the cell and the test mass returning the probe beam is used to minimize the stand-off distance from the cell to the test mass. Any optical setup for the probe beam detection would require some space and might influence the measurements, while the minimization of the distance between the cell and the test mass is important for increasing sensitivity to the interactions at the region of small interaction ranges. There are three possible variations in the setup to search for the spin-dependent interactions: (1) the test mass is arranged to move forward and backward along the $\hat{z}$ axis, (2) right and left along the $\hat{x}$ axis with a velocity $\vec{v}$; (3) to revolve clockwise or counterclockwise around the $\hat{z}$ with a constant angular frequency. The direction of the spin $\hat{\sigma}_2$ in the polarized test mass would be chosen along the $\hat{x}$, $\hat{y}$ or $\hat{z}$ axes to probe different interactions.

For an unpolarized test mass, we assume to use a nonmagnetic bismuth germanate insulator (Bi$_{4}$Ge$_{3}$O$_{12}$, or BGO) with the high nucleon density ($7.13$ g/cm$^{3}$ = $4.3\times 10^{24}$ nucleons/cm$^{3}$) which has been previously used~\cite{Tullney2013}. In this paper, we only theoretically calculated the spin-dependent interactions between the polarized electrons in the vapor cell and the unpolarized nucleons in the BGO test mass. For a polarized mass, Dy$_{6}$Fe$_{23}$ and HoFe$_{3}$ with the electron spin density $1.6\times 10^{22}$ spins/cm$^3$~\cite{Chui1993,Ni1994} and layers of ferromagnet Alnico 5 and SmCo$_5$ with the electron spin density $(3.66\pm0.08)\times 10^{22}$ spins/cm$^3$~\cite{Heckel2008, Hoedl2011, Heckel2013, Terrano2015} were used. Here we consider a dysprosium iron garnet (Dy$_{3}^{3+}$Fe$_{2}^{3+}$Fe$_{3}^{3+}$O$_{12}$, or DyIG) with the spin density of $4\times 10^{20}$~spins/cm$^{3}$ and zero magnetization at the critical temperature $T_c = 226$ K, which has been previously investigated~\cite{Leslie2014}. Another choice is terbium iron garnet (TbIG) for its higher critical temperature, $T_c = 266$ K, while the spin density is reduced by a factor of two~\cite{Leslie2014}.

\begin{table}[t]
\begin{center}
\begin{tabular}{ l | r}
\hline
\hline
parameter & value\\
\hline
\hline
cell radius & 3 cm \\
cell window thickness & 0.1 cm\\
gap between Rb gas and mass & 1 cm\\
magnetic field sensitivity & 10 fT/$\sqrt{\text{Hz}}$\\
test mass dimension & $5\times 5 \times 5$ cm$^{3}$\\
BGO density & $7.13$ g/cm$^{3}$ \\
BGO nucleon density & $4.3\times 10^{24}$ /cm$^{3}$\\
DyIG spin density & $4\times 10^{20}$/cm$^{3}$\\
%maximum velocity & 3 cm/s\\
modulation amplitude & 0.5 cm\\
modulation frequency & 10 Hz\\
rotation frequency & 10 Hz\\
\hline
\end{tabular}
\end{center}
\caption{Experimental parameters used for the estimation of the sensitivity of the proposed experiment to the spin-dependent interactions.}
\label{tab:parameter}
\end{table}
\begin{figure}[h]
\includegraphics[width=0.45\textwidth]{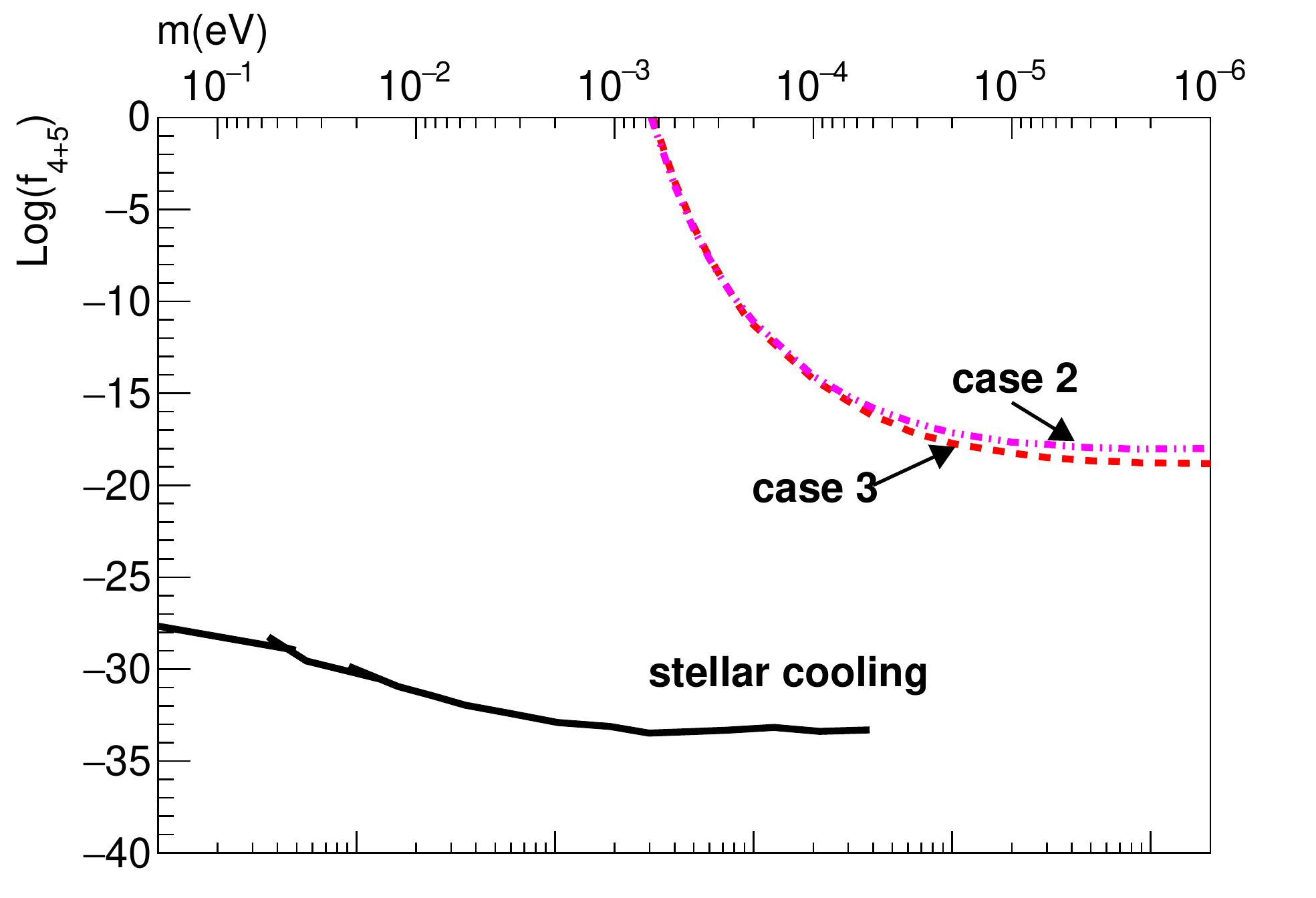}
\includegraphics[width=0.45\textwidth]{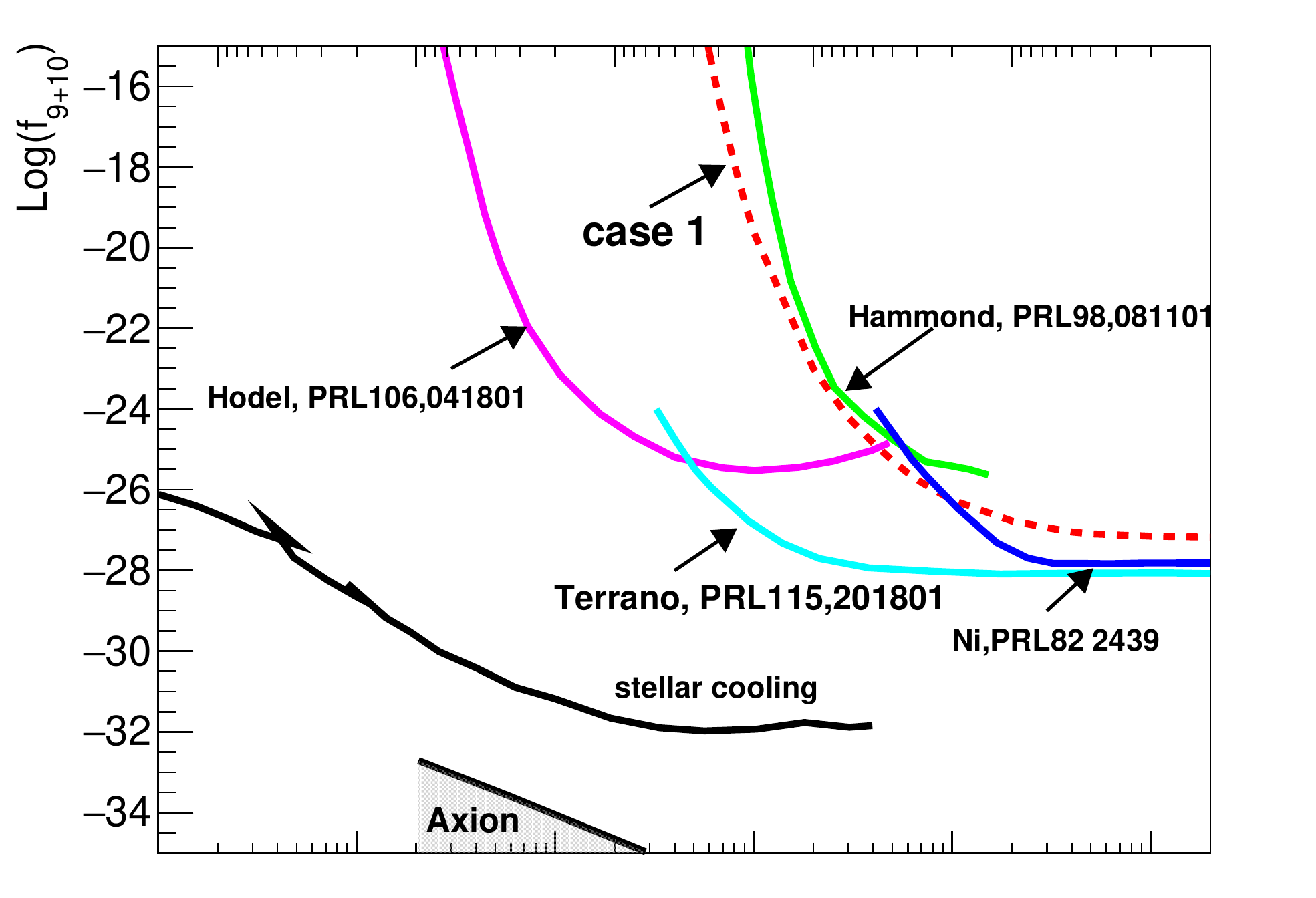}
\includegraphics[width=0.45\textwidth]{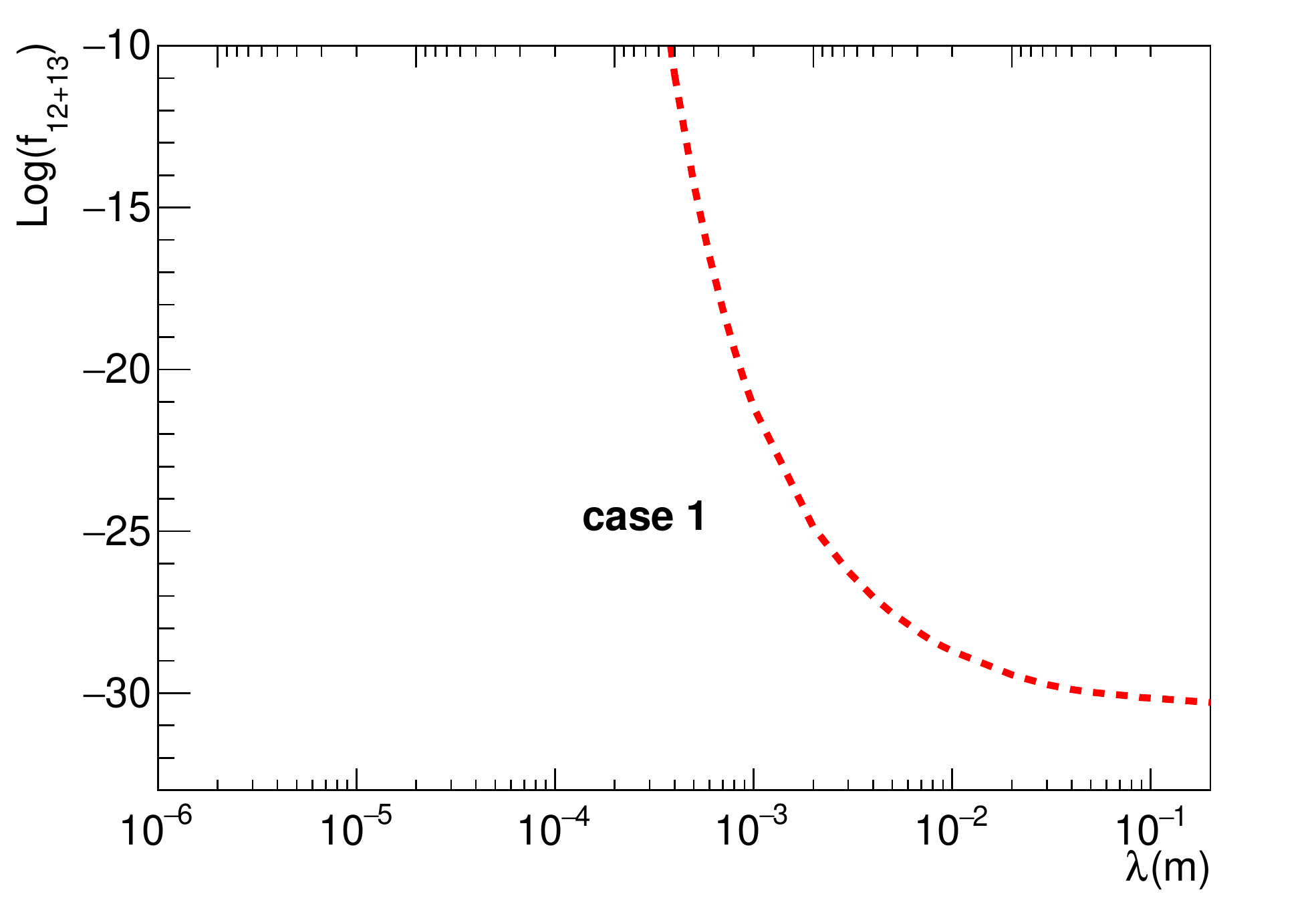}
\caption{ The constraints of the coupling strength to the interactions, from top to bottom, $V_{4+5}$, $V_{9+10}$, $V_{12+13}$~\cite{Dobrescu2006}, as a function of the interaction range (bottom axes) and the ALP mass (top axes). The dashed curves are the estimated sensitivity of the proposed experiment to the interactions between the polarized Rb electrons and the unpolarized BGO~\cite{Tullney2013} test mass for one second measurement period. The solid curves are current limits. The axion coupling strength and range~\cite{Moody1984, Leslie2014} is shown in $V_{9+10}$. The constraints from the stellar cooling for the axion together with short-range gravity experiments with unpolarized masses are also shown in $V_{4+5}$ and $V_{9+10}$~\cite{Raffelt:2012}.}
\label{fig:sensitivity}
\end{figure}
As listed in Table~\ref{tab:schematic}, there are nine combinations between $\hat{\sigma}_1$, $\hat{\sigma}_2$ and $\vec{v}$, corresponding to different interactions. In order to estimate the experimental sensitivity we assume the experimental parameters listed in Table~\ref{tab:parameter}. The volume of the test mass is chosen to be $5\times 5 \times 5$ cm$^3$ to match approximately the diameter of the vapor cell. The realistic closest distance between the Rb spins and the test mass is about 1~cm because of the cell wall thickness and the required heat insulation. The test mass can be held at different temperatures while the SERF cell requires 150$^\circ$C.  The test mass position can be moved using a linear actuator with a modulation of $A\sin(2\pi f t)$, where $A$ is set at 0.5~cm (the half of the maximum distance that the mass moves), $f$ is the frequency of the modulation, and $t$ is the time. A modulation frequency of 10~Hz will be used to avoid the 1/$f$ noise usually present in the SERF magnetometer signal. If the modulation is along the $\hat{z}$ axis, the maximum velocity  of 31.4~cm/s is achieved when the mass is 1.5~cm away from the Rb spins. This distance will determine the sensitivity to the velocity-dependent interactions.
%; otherwise, the minimum distance between the test mass and the Rb spins is limited by the cell wall thickness and the heat insulation.
If the modulation direction is parallel to the $\hat{x}$ axis, the mass can slide along the minimum distance 1 cm, then the maximum velocity will be when the center of the mass is near the axis of the Rb cell. A frequency of 10~Hz and an amplitude of 0.5~cm can be chosen to have the maximum velocity of 31.4~cm/s. The mass can be also rotated around the $\hat{z}$ axis with a motor at a frequency of 10~Hz. In this case, the minimum distance between the test mass and the Rb spins is also determined by the cell wall thickness and the heat insulation. In case of rotation, frequency can be in principle increased if the 1/$f$ noise is still significant. 

\begin{figure}[h]
\includegraphics[width=0.45\textwidth]{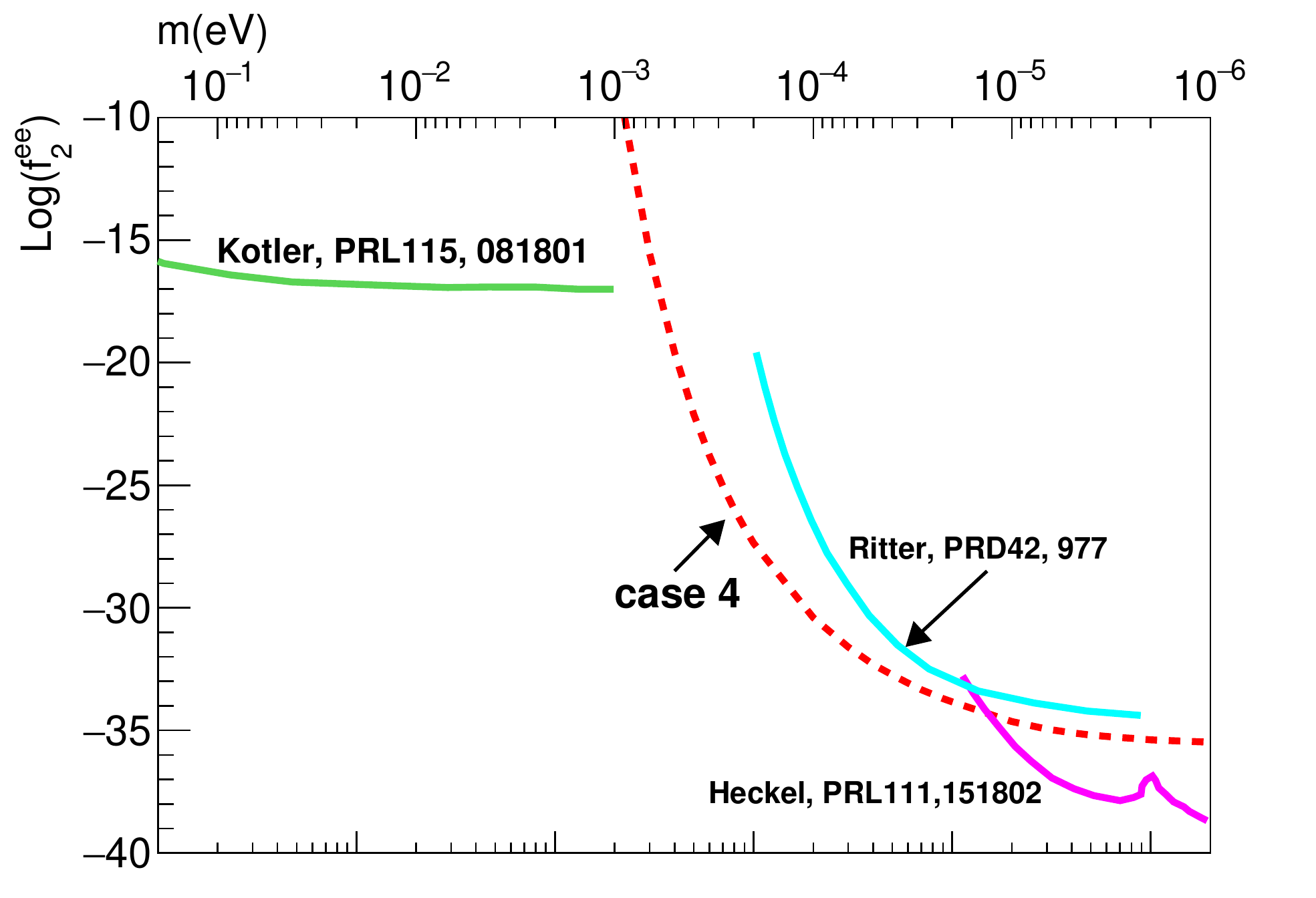}
\includegraphics[width=0.45\textwidth]{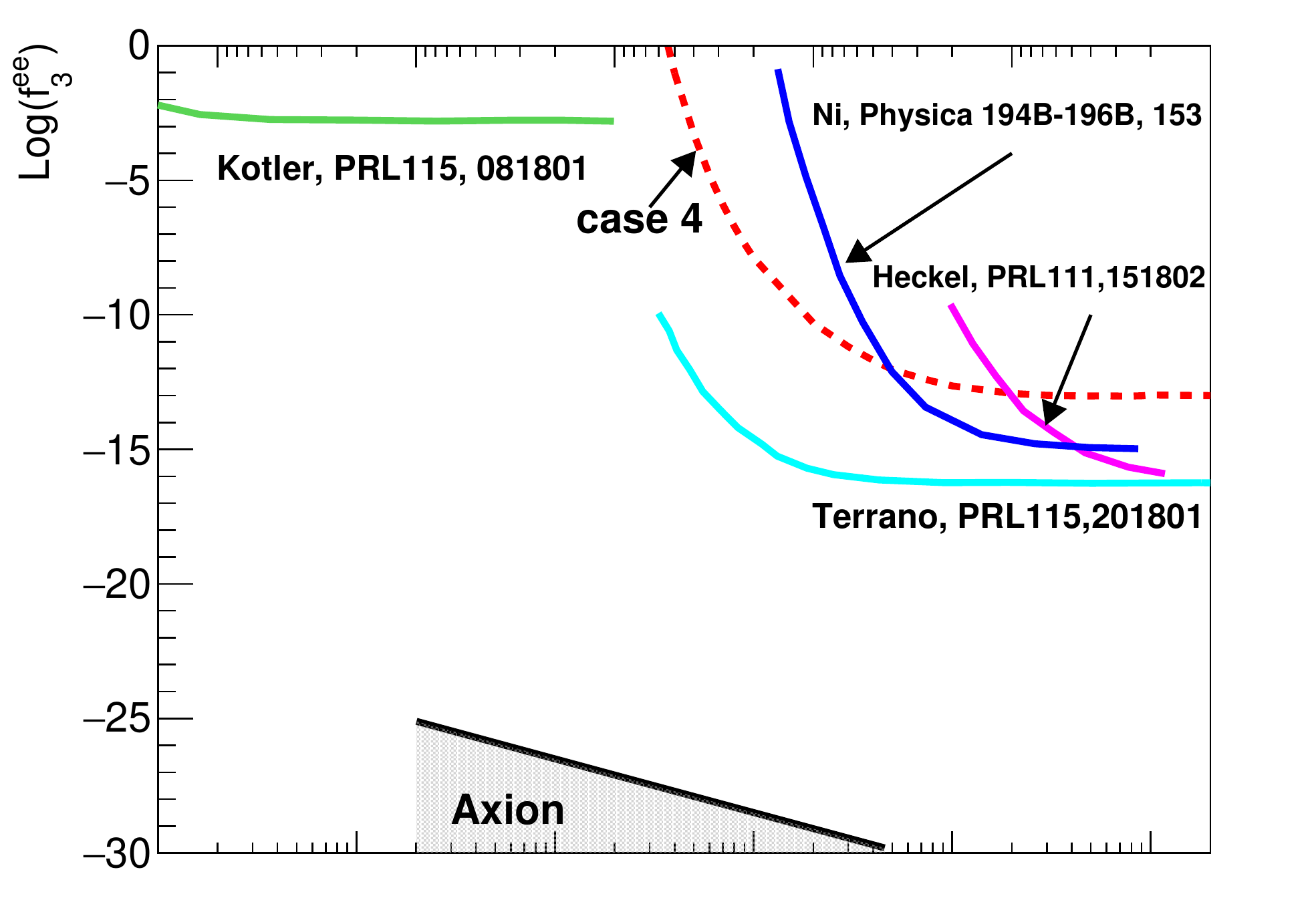}
\includegraphics[width=0.45\textwidth]{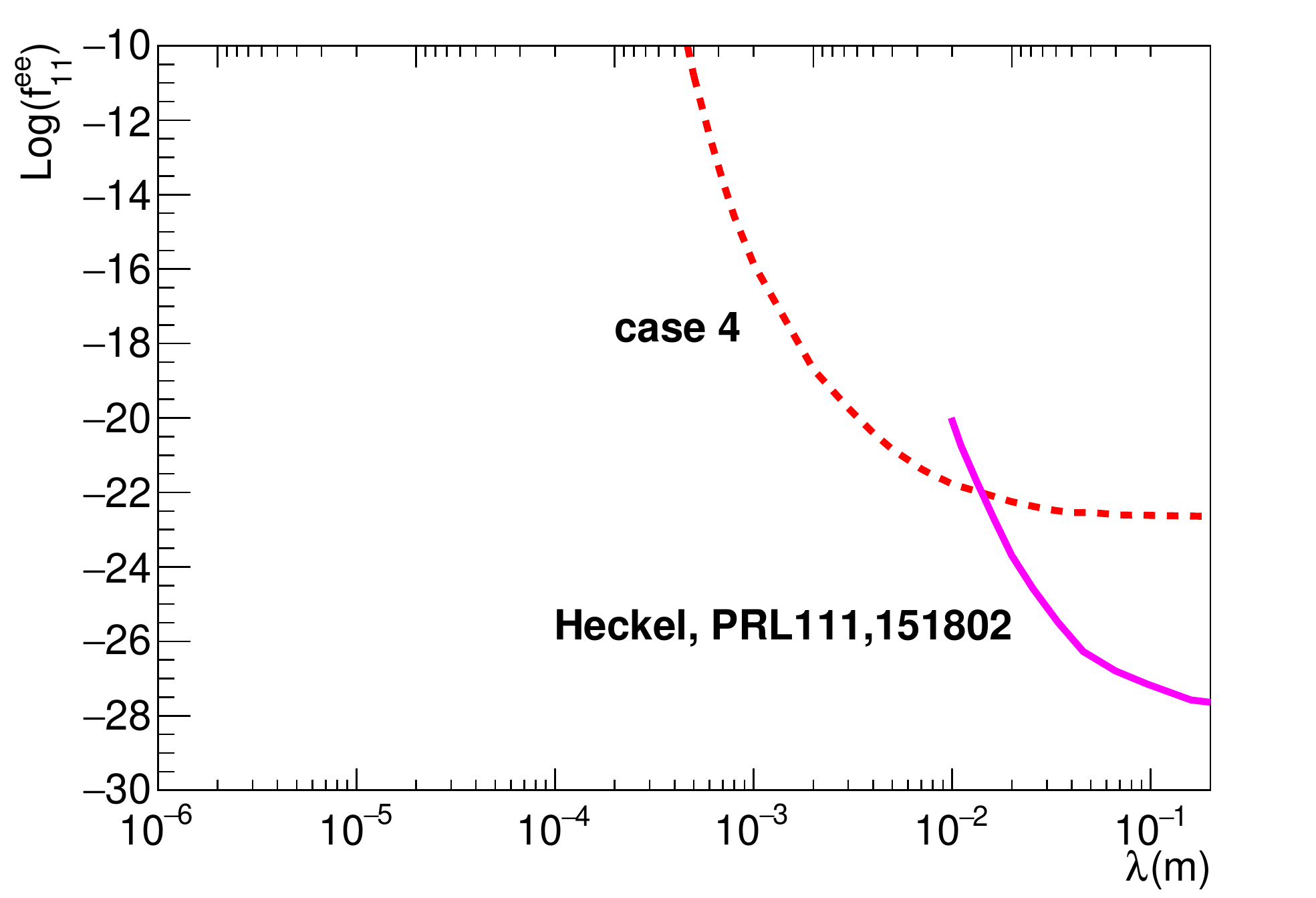}
\caption{The constraints of the coupling strength to the interactions, from top to bottom, $V_2$, $V_3$, $V_{11}$~\cite{Dobrescu2006}, as a function of the interaction range (bottom axes) and the ALP mass (top axes). The dashed curves are the estimated sensitivity of the proposed experiment to the interactions between the polarized Rb electrons and the polarized DyIG~\cite{Leslie2014} test mass for one second measurement period. The solid curves are current limits. The axion coupling strength and range~\cite{Moody1984, Leslie2014} is shown in $V_{3}$. }
\label{fig:sensitivity-2}
\end{figure}

\begin{figure*}
\begin{minipage}[b]{0.45\textwidth}
\centering
\includegraphics[width=1\textwidth]{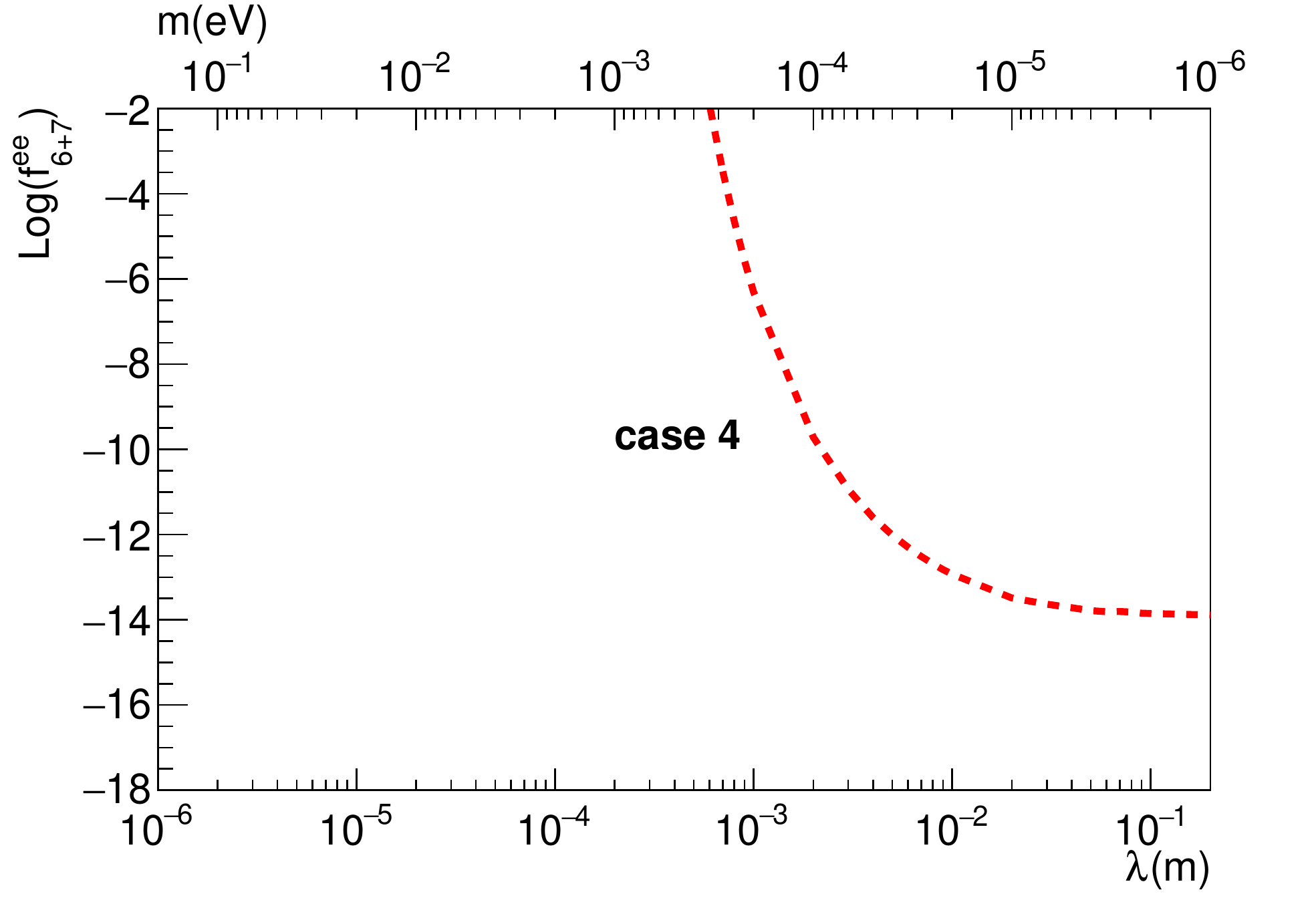}
\includegraphics[width=1\textwidth]{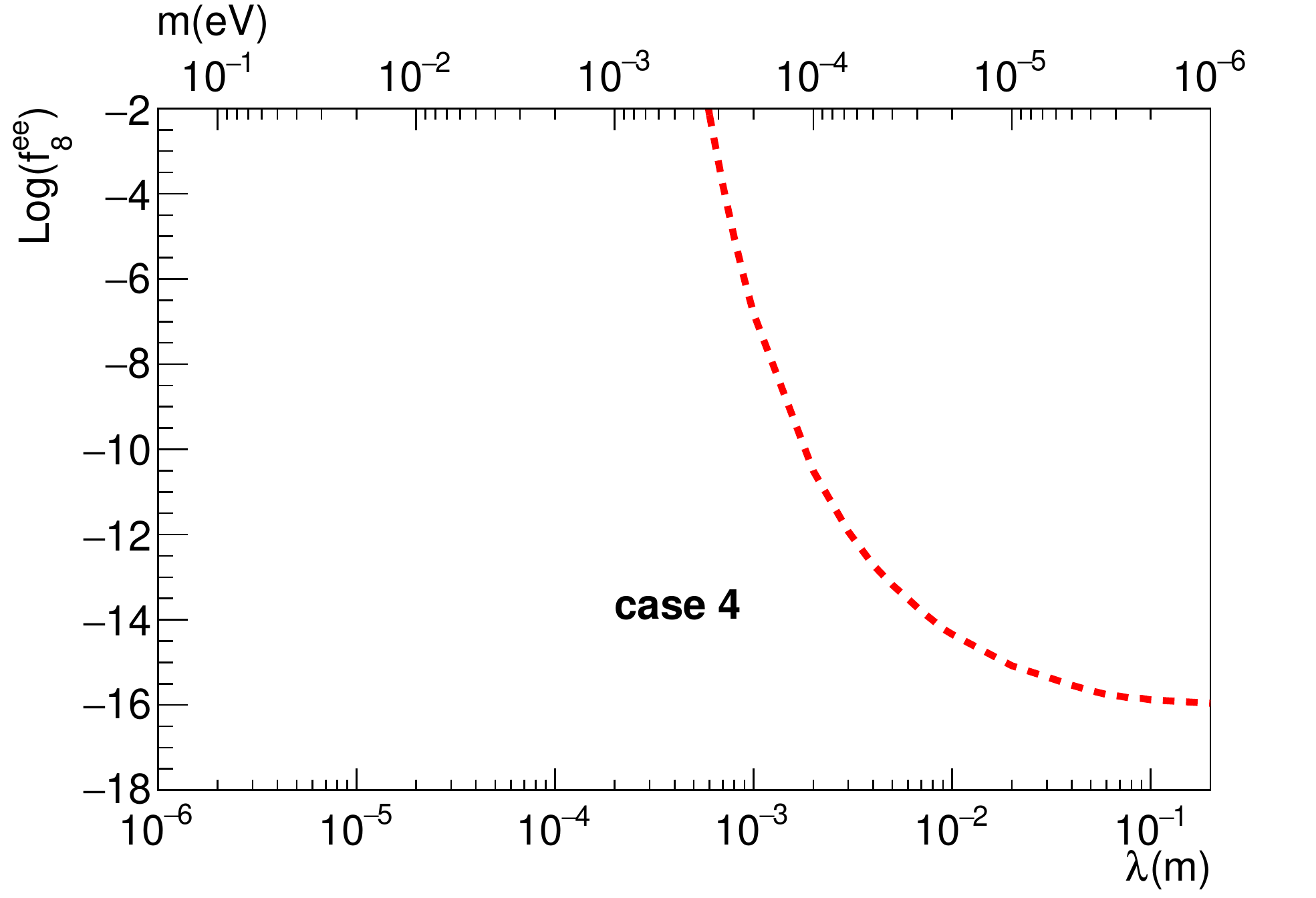}
\end{minipage}
\begin{minipage}[b]{0.45\textwidth}
\includegraphics[width=1\textwidth]{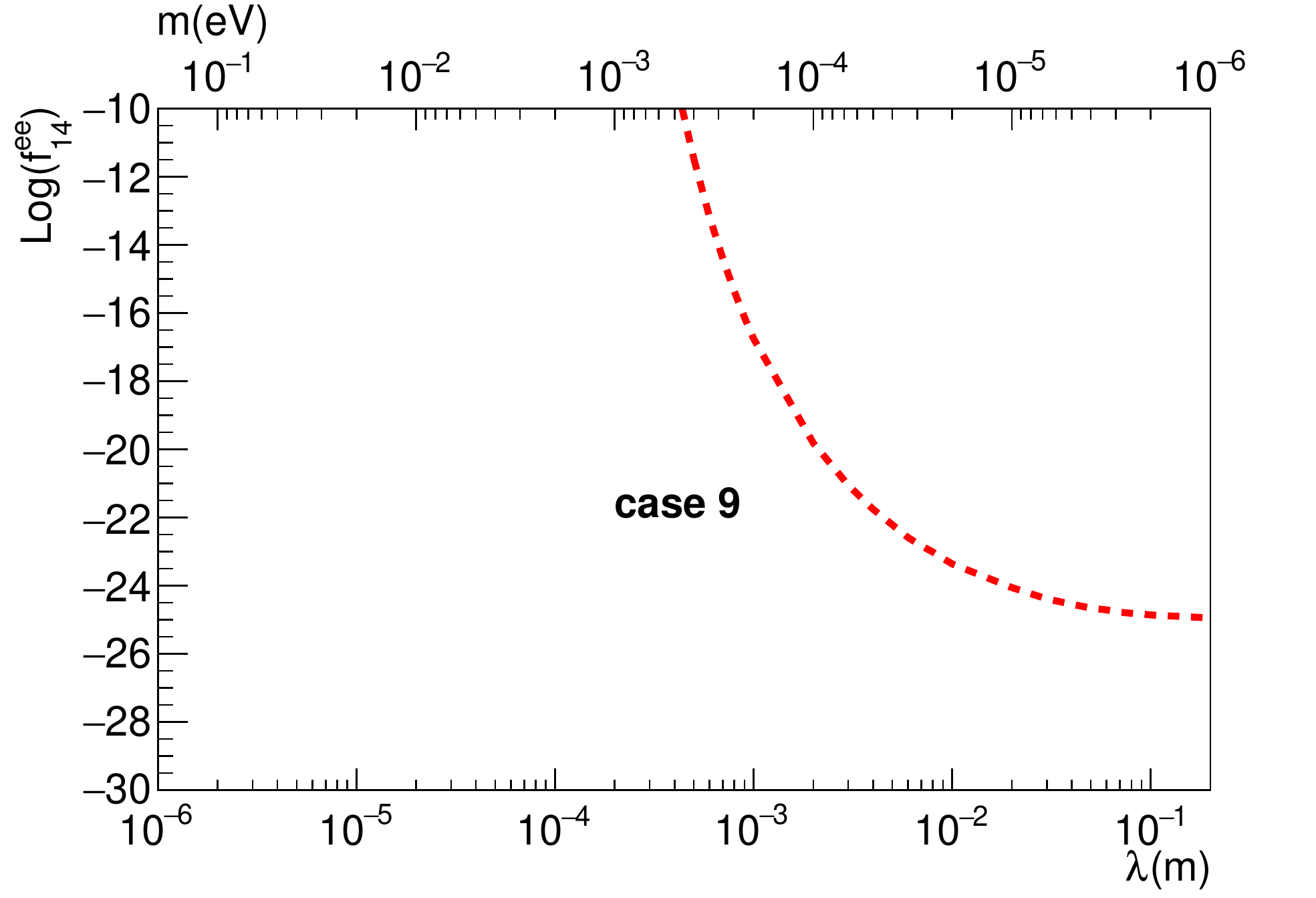}
\includegraphics[width=1\textwidth]{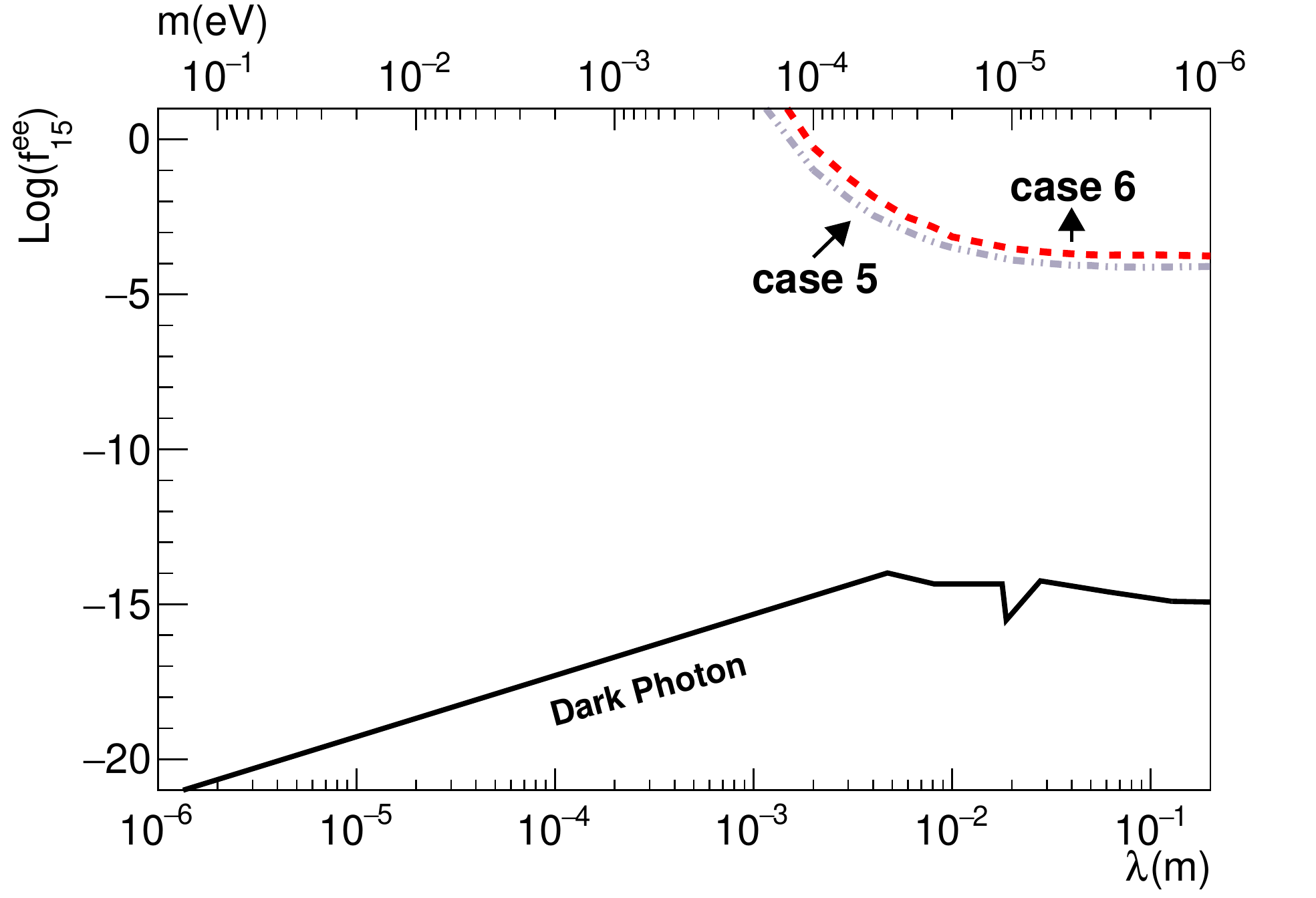}
\end{minipage}
\centering
\includegraphics[width=0.45\textwidth]{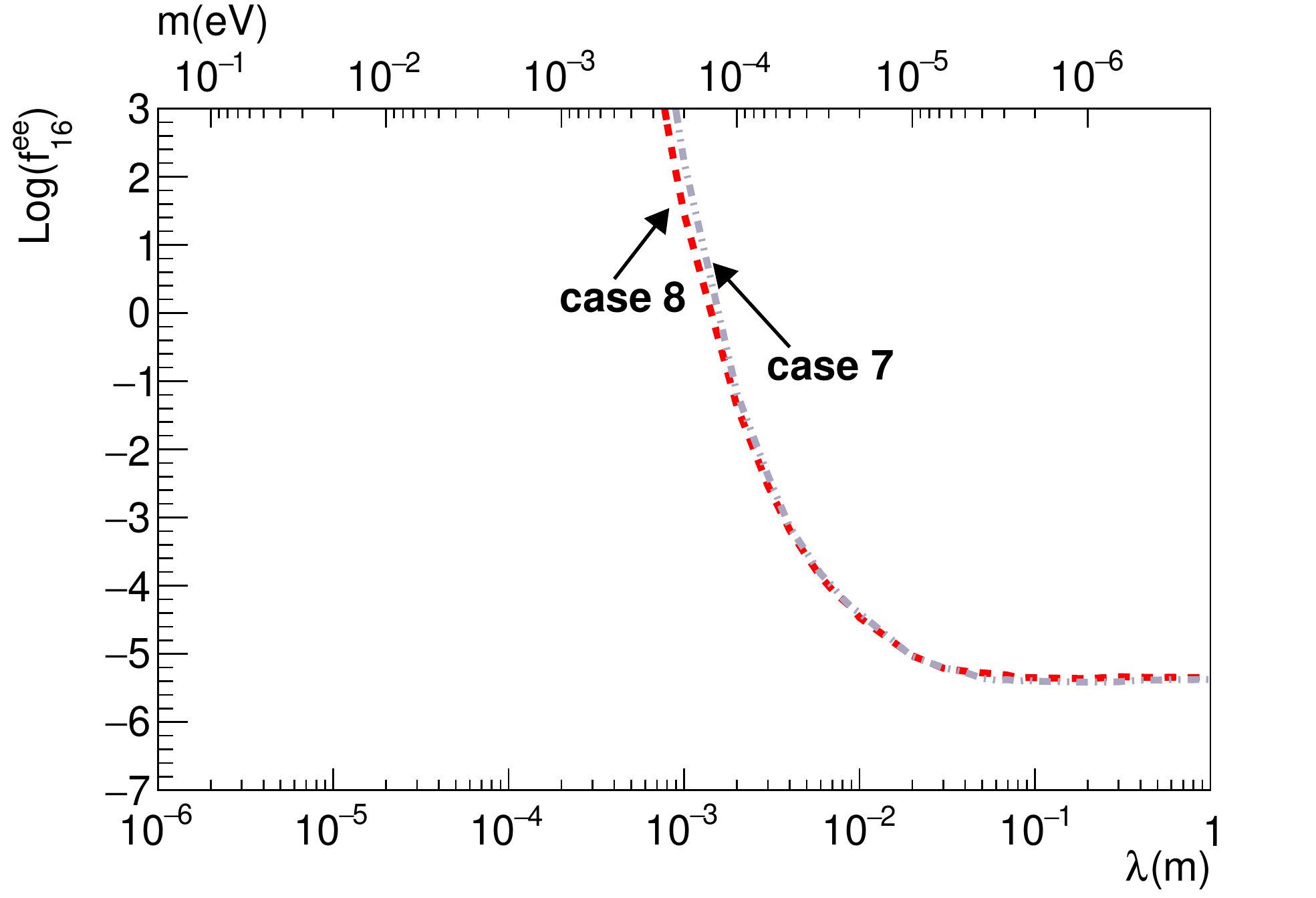}
%\subfigure[$f_{14}$]{}}\qquad
%\subfigure[$f_{15}$]{\includegraphics[width=0.46\textwidth]{./sensitivity_15.pdf}}\qquad
%\subfigure[$f_{16}$]{\includegraphics[width=0.46\textwidth]{./sensitivity_16.pdf}}
\caption{The constraints of the coupling strength to the interactions, $V_{6+7}$, $V_8$, $V_{14}$, $V_{15}$ and $V_{16}$~\cite{Dobrescu2006}, as a function of the interaction range (bottom axes) and the ALP mass (top axes). The dashed curves are the estimated sensitivity of the proposed experiment to the interactions between the polarized Rb electrons and the polarized DyIG~\cite{Leslie2014} test mass for one second measurement period. The solid curves are current limits. The constraints from the dark photon~\cite{Essig:2013} are also shown in $V_{15}$.} 
\label{fig:sensitivity-3}
\end{figure*}

In order to estimate the sensitivity of the proposed experiments to the spin-dependent interactions, we applied the Monte Carlo method to average the interaction potentials given by Eq.~\ref{eq:v45}-\ref{eq:v16} between the test mass (BGO for unpolarized mass and DyIG for polarized mass) and the Rb spins. First, random points were generated in the volume of the test mass and the Rb cell. Then, the interaction range was assumed to calculate the potential between two randomly generated points. Next, all the contributions to the potential were summed and normalized to give the average potential for the densities of particles. Finally, the coupling strength for a typical magnetic field sensitivity ($\sim$10~fT per second, the demonstrated sensitivity of LANL pancake magnetometer, which corresponds to the energy shift of $1.8\times10^{-18}$ eV for Rb atoms) was derived. Fig.~\ref{fig:sensitivity} shows the sensitivity to the interactions between unpolarized nucleons in BGO and polarized Rb electrons: $V_{4+5}$, $V_{9+10}$ and $V_{12+13}$. It can be seen that for the $V_{9+10}$ potential which value was constrained by the experiments with torsion pendulum~\cite{Hoedl2011,Terrano2015} the SERF magnetometer experiment does not offer any improvement. However, the other two interactions for polarized electrons have not been constrained by experiments, and therefore our proposed experiments will be of great value. Fig.~\ref{fig:sensitivity-2} shows the sensitivity to interactions between two polarized electrons in DyIG and the cell which are independent of the velocity, $V_{2}$, $V_{3}$ and $V_{11}$. The present experimental constraints of these interactions for two polarized electrons were obtained from torsion pendulum~\cite{Ritter1993, Heckel2013,Terrano2015} and paramagnetic salt~\cite{Ni1999} measurements. The estimation shows that  the SERF magnetometer has no advantage for $V_{3}$ but can be sensitive to the new phase space of $V_{2}$ and $V_{11}$. Fig.~\ref{fig:sensitivity-3} shows the sensitivity to interactions between two polarized electrons dependent on the velocity, $V_{6+7}$, $V_{8}$, $V_{14}$, $V_{15}$ and $V_{16}$. Because there are no experimental constraints on these interactions between two polarized electrons so far, this experiment could set new limits on the coupling strength of these interactions. In principle, the experimental sensitivity can be enhanced by repeating measurement; for $N$ times measurement, the sensitivity will be enhanced by $1/\sqrt{N}$ until systematic errors become dominant. 

In order to compare our estimated sensitivity with the existing constraints on the axion, we rescaled the predicted coupling strength and the range of the axion for the case of a spin-0 interaction in Ref.~\cite{Moody1984} to $f_3$ and $f_{9+10}$~\cite{Leslie2014}. The 10 meV cutoff~\cite{Rosenberg:2000} is the limit from SN1987a~\cite{Engel:1990}. On other hand, the pseudoscalar coupling of electrons is strongly constrained by the stellar cooling for the axion~\cite{Raffelt:1995}. Together with short-range gravity experiments with unpolarized masses for the scalar coupling of electrons, the more strong constraints for $V_{4+5}$ and $V_{9+10}$ can be derived~\cite{Raffelt:2012}, shown in Fig.~\ref{fig:sensitivity} and Fig.~\ref{fig:sensitivity-2}. The dark photon~\cite{Essig:2013},  as a vector coupling boson, strongly constrains the vector coupling of electrons. The limit from the dark photon~\cite{Essig:2013} can provide the constraints for $V_{15}$ shown in Fig.~\ref{fig:sensitivity-3}, which is only dominated by the square of the vector coupling of electrons~\cite{Leslie2014}. This relation can be derived by comparing the Eq. 4 of~\cite{Essig:2013} and Eq. 5.18 of~\cite{Dobrescu2006}.

Systematic errors could be induced from the magnetic susceptibility of the test masses, which fluctuates with temperature and the magnetic field drifts. One challenge is precise control of temperature. The Rb cell needs to be kept at $\sim$150$^\circ$C while the test mass at room temperature for BGO and low temperature for DyIG or TbIG at 226~K and 266~K, respectively. The temperature difference between the Rb cell and the test mass may change the magnetic susceptibility and cause additional magnetization on the test mass. An alternative setup is to put the Rb cell in a magnetic shield so that the cell will not feel the magnetic effect of the test mass while the test mass can be kept at the room temperature though a recent study implies a possible interaction between the magnetic shielding and exotic spin-dependent interactions~\cite{Kimball:2016}; however, this setup will increase the distance between the Rb spins and the test mass. Another challenge is to make the distance between the Rb cell and the test mass as short as possible in order to maximize the sensitivity to the spin-dependent interactions at smaller interaction ranges, which is currently limited by the cell wall thickness and the heat insulation.

Apart from controlling temperature and magnetic field, one way to reduce the systematic error is to compare the signals between the different states of the modulation. For the interactions which have only one term of the velocity, the interaction-induced signal will be in opposite sign while the orientation of the velocity is flipped. On the other hand, the systematic error due to the magnetic susceptibility and the magnetization of the test mass will be the same for opposite velocity orientations. Therefore, for $V_{4+5}, V_{6+7}, V_{12+13}, V_{14}$ and $V_{15}$, the signal comparison between the states of the opposite velocity orientation could suppress the systematic error. The method of the SERF magnetometer should be mostly sensitive to these interactions. In case of the interactions which are dependent on the vector along the direction between two interacting particles, including $V_{9+10}$, $V_{11}$ and $V_{16}$, the sign of the interactions will change while the test mass is put at the opposite ends of the Rb cell. One possible way to module the interaction is to install one test mass at one end of the Rb cell and another test mass at the opposite end of the Rb cell. The residual systematic error may be due to the magnetic field gradient across the Rb cell. For interactions which are only dependent on two spins (or two velocities, two distance vectors), including $V_{2}, V_{3}$ and $V_{8}$, there is no proper way to modulate the test mass and cancel out the systematic error. The only way is to flip the spin orientation of the Rb electrons or the polarized test mass. However, flipping spin orientation could cause a large systematic error because the magnetization is along the spin orientation as well. Therefore, the method of the SERF magnetometer will be less sensitive to these interactions, including $V_{2}, V_{3}$ and $V_{8}$, unless the temperature as well as the magnetic susceptibility can be well controlled.

The setup of the test mass described in this paper can be also applied to other systems with different test masses to test spin-dependent and velocity-dependent interactions: for example, polarized atoms~\cite{Youdin1996,Chu2013,Bulatowicz2013,Tullney2013} for polarized nucleons, paramagnetic insulator~\cite{Chu2015} for polarized electrons. 

In conclusion, we proposed SERF magnetometer-based experimental methods to search for the exotic spin-dependent interactions for polarized electrons. Our detailed calculations of the projected experimental sensitivity showed that the experiments are sensitive to the interactions, especially at the interaction range of $10^{-2}$ to $10^{-4}$ m, most of which are not experimentally constrained. The possible experimental setups of the movement direction and the spin orientation of the test mass shown in Table~\ref{tab:schematic} play a key role in probing the different spin-dependent interactions. We also described challenges in improving experimental sensitivity at small interaction ranges and reducing possible systematic errors due to variation of the magnetic susceptibility of a test mass. One way to suppress the systematic errors is to module the interaction-induced signals by modulating the test mass. 

The authors thank H. Gao, J. Long and W. M. Snow for useful discussions.  %This work is supported by the U. S. DOE through the LANL/LDRD program.
%\newpage

\end{document}